\documentclass[aps,twocolumn,superscriptaddress,floatfix]{revtex4-1}
\usepackage{graphicx,amsfonts,amssymb,amsmath,mathrsfs,hyperref,bm}

\newif\ifhyper
\hypertrue
\ifhyper
\hypersetup{
   citecolor = {red},
   colorlinks = {true}, 
   linkcolor = {blue},
   urlcolor = {blue} 
}
\fi

\def\be{\begin{equation}}
\def\ee{\end{equation}}
\def\bea{\begin{eqnarray}}
\def\eea{\end{eqnarray}}

\newcommand{\ket}[1]{|#1\rangle}
\newcommand{\bra}[1]{\langle #1|}
\newcommand{\braket}[2]{\langle #1|#2\rangle}

\newcommand{\expectval}[1]{\langle #1\rangle}

\begin{document}

\title{Spin-$\frac{1}{2}$ Heisenberg antiferromagnet on the star lattice: \\ Competing valence-bond-solid phases studied by means of tensor networks}

\author{Saeed S. Jahromi}
\email{jahromi@physics.sharif.edu}
\affiliation{Department of Physics, Sharif University of Technology, Tehran 14588-89694, Iran}

\author{Rom\'an Or\'us}
\affiliation{Institute of Physics, Johannes Gutenberg University, 55099 Mainz, Germany}
\affiliation{Donostia International Physics Center, Paseo Manuel de Lardizabal 4, E-20018 San Sebasti\'an, Spain}
\affiliation{Ikerbasque Foundation for Science, Maria Diaz de Haro 3, E-48013 Bilbao, Spain}

\begin{abstract}
Using the  infinite Projected Entangled Pair States (iPEPS) algorithm, we study the ground-state properties of the spin-$1/2$ quantum Heisenberg antiferromagnet on the star lattice in the thermodynamic limit. By analyzing the ground-state energy of the two inequivalent bonds of the lattice in different unit-cell structures, we identify two competing Valence-Bond-Solid (VBS) phases for different antiferromagnetic Heisenberg exchange couplings. More precisely, we observe (i) a VBS state which respects the full symmetries of the Hamiltonian, and (ii) a resonating VBS state which, in contrast to previous predictions, has a six-site unit-cell order and breaks $C_3$ symmetry. We also studied the ground-state phase diagram by measuring  the ground-state fidelity and energy derivatives, and further confirmed the continuous nature of the quantum phase transition in the system. Moreover, an analysis of the isotropic point shows that its ground state is also a VBS as in (i), which is as well in contrast with previous predictions. 

\end{abstract}

\maketitle

\section{Introduction.} 

The interplay between antiferromagnetic (AF) Heisenberg interactions and geometric frustration in 2d quantum magnets has been known to be the root of many exotic phases of matter, ranging from Valence-Bond-Solids (VBS) \cite{Evenbly2010,Hastings2000,Marston1991,Singh2008} to quantum spin liquids (QSL) \cite{Yang1993,Yan2011,Wang2006,Ran2007,Depenbrock2012} with or without topological order \cite{Wen1995,Kitaev2006,Jahromi2013,Jahromi2013a,Jahromi2016,Jahromi2017}. The Shastry-Sutherland \cite{SriramShastry1981,Corboz2014} model and the spin-$S$ kagome Heisenberg antiferromagnet  \cite{Picot2016,Evenbly2010,Hastings2000,Marston1991,Singh2008} are two well known examples of frustrated systems of  experimental relevance \cite{Hiroi2001,Okamoto2007,MatthewP.Shores2005}, and for which the true nature of their ground states is still under debate even after tremendous analytical and numerical efforts \cite{Corboz2014}.

The antiferromagnetic Heisenberg model on the star lattice (AFHS, see Fig.~\ref{Fig:PhaseDiag}) \cite{Richter2004} is another interesting example of a 2d quantum system with potentially-exotic ground states. Due to geometric frustration and 
lower coordination number, the AFHS model can impose stronger quantum fluctuations on the two inequivalent bonds of the star lattice. This makes the star lattice even more resourceful than the kagome lattice \cite{Picot2016} when it comes to hosting exotic phases of matter. Previous studies found that the ground state of spin-$\frac{1}{2}$ models on the star lattice can host exact chiral QSL states with non-Abelian fractional statistics \cite{Yao2007,Kells2010,Dusuel2008}, VBS ground-states \cite{Richter2004,Misguich2007,Yang2010,Choy2009}, trimerized spin-$1$ states \cite{Ran2018} and topological order in different QSL phases \cite{Huang2013}. Besides, the ground states of the spin-$1$ bilinear-biquadratic model on the star lattice are given by different phases among which the QSL and ferroquadrupolar states are of potential interests \cite{Lee2018}.

In spite of considerable progress towards the characterization of quantum paramagnetic phases on the star lattice \cite{Richter2004,Misguich2007,Yang2010,Choy2009}, a general understanding of the interplay between competing phases upon the variation of the spin interactions is still lacking. Particularly, the true nature of competing ground states of the spin-$1/2$ AFHS for different Heisenberg exchange couplings is not fully understood. Previous studies with exact diagonalization (ED) on finite clusters \cite{Richter2004,Misguich2007} and mean-field results based on Gutzwiller projected wave functions \cite{Yang2010} predicted that the spin-$1/2$ AFHS hosts two competing VBS states for different values of the couplings. More specifically, a VBS state which fully respects the symmetries of the Hamiltonian, and a VBS state with $\sqrt{3}\times\sqrt{3}$ ordering \cite{Yang2010}. However, it is not yet clear if this ordering persists in the thermodynamic limit or if it is an artifact due to finite-size effects. Additionally, examples of frustrated spin systems with a star lattice structure have recently been discovered in Iron Acetate \cite{Zheng2007}. Thus, investigating the true nature of these competing VBS states becomes of crucial importance.

Motivated by this, we use an improved version of the powerful infinite Projected Entangled-Pair State algorithm (iPEPS)  \cite{Verstraete2004,Verstraete2006,Vidal2007}, adapted for nearest-neighbor local Hamiltonians on the triangle-honeycomb lattice \cite{Jahromi2018}, to study the spin-$1/2$ AFHS model in the thermodynamic limit. The iPEPS method does not suffer from the infamous sign problem for frustrated spin models and provides a variational ground state energy and wave function. Using this algorithm we are able to extract remarkably small energy differences on topologically equivalent bonds of the lattice, and reveal the stable ordering of the underlying VBS states in the thermodynamic limit. We further map out the full phase diagram of the AFHS model in the thermodynamic limit by measuring the ground-state fidelity \cite{Zhou2008,Jahromi2018} as well as energy derivatives, in turn capturing the continuous nature of the quantum phase transition in the system.

The paper is organized as follows. In Sec.~\ref{sec:modelmethod} we introduce the AFH model on the star lattice and briefly discuss the details of the iPEPS machinery we used for evaluating the ground-state of the system. Next, in Sec.~\ref{sec:vbsphases} we characterize the different VBS phases which emerge in different Heisenberg exchange couplings. We study the two-pint correlations in VBS phases of the AFHS model in Sec.~\ref{sec:correlators} and investigate the quantum phase transition in the system in  Sec.~\ref{sec:qpt}. Finally Sec.~\ref{sec:conclude} is devoted to conclusion and outlook for future studies.

\section{Model and Method} 
\label{sec:modelmethod}  
  
\begin{figure}
\centerline{\includegraphics[width=\columnwidth]{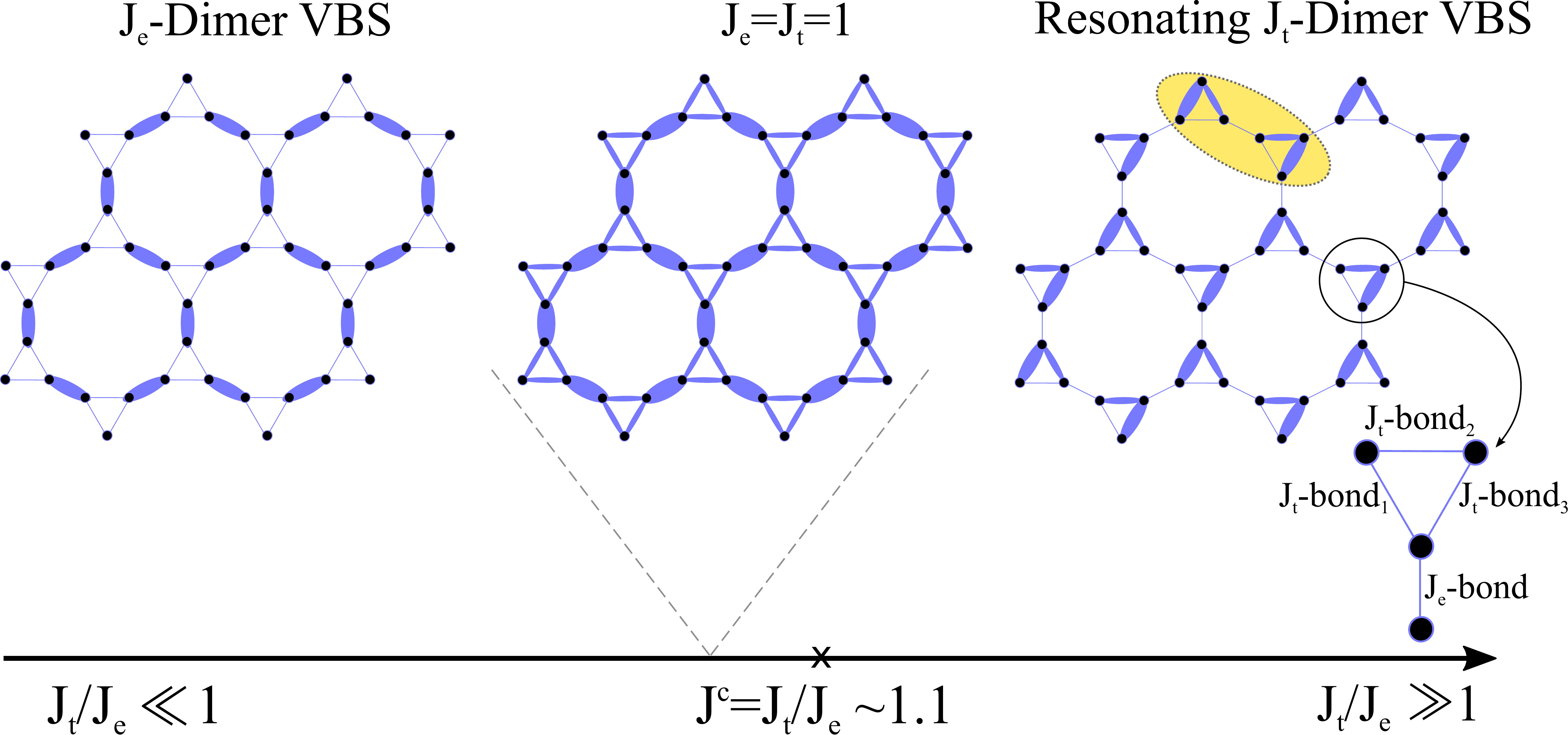}}
\caption{(Color online) Phase diagram of the AFH model on the star lattice. (left) The ground state of the system in the large-$J_e$ limit is a $J_e$-dimer VBS state which fully respects the symmetries of the Hamiltonian. (center) At the isotropic point, $J_e=J_t=1$, the system is still in a $J_e$-dimer VBS state with stronger bond energies on the triangles.  (right) The ground-state at large-$J_t$ is a resonating $J_t$-dimer VBS, which breaks the local $C_3$ symmetry of the triangles, and with a six-site unit cell order (yellow region). There is a continuous phase transition between the VBS phases at $J^c=J_t/J_e\approx 1.1$.}
\label{Fig:PhaseDiag}
\end{figure}

The spin-$1/2$ antiferromagnetic Heisenberg model on the star lattice is given by 
\be
\label{eq:H_AFHS}
H = J_e \sum_{\langle i j \rangle \in e} \mathbf{S}_i \cdot \mathbf{S}_j + J_t \sum_{\langle i j \rangle \in t} \mathbf{S}_i \cdot \mathbf{S}_j, 
\ee
where the first sum runs over edges, $e$, connecting the triangles, and the second sum runs over the links, $t$, of the triangles of the star lattice (see Fig.~\ref{Fig:PhaseDiag}). $J_e$ and $J_t$ are the antiferromagnetic (AF) Heisenberg exchange couplings and $\mathbf{S}_i$ is the spin operator at lattice site $i$. As discussed above, this model exhibits different competing VBS states for different values of the exchange couplings $J_e$ and $J_t$ \cite{Richter2004,Yang2010}. 

In order to study the ground states of the system for different coupling regimes, we use the tensor network method (TN) \cite{Orus2014} based on the iPEPS technique \cite{Verstraete2004,Verstraete2006,Vidal2007,Corboz2010}. The method uses a tensor network variational ansatz to approximate the 2d ground states in the thermodynamic limit. Here, we use an improved version of iPEPS that we developed for the triangle-honeycomb lattice \cite{Jahromi2018} for arbitrary unit cell sizes, and approximate the ground state of the system via imaginary time evolution. As discussed in Appendix.~\ref{app:ipeps}, we use two different tensor network setups to describe the states on the star lattice. Specifically, setup A (B) is designed to capture the VBS order in large-$J_e$ ($J_t$) limit, in the thermodynamic limit. The control parameters of our algorithm are the PEPS bond dimension $D$, which controls the maximum amount of entanglement handled by the wavefunction, and the environment bond dimension $\chi$, which controls the accuracy of the approximations when contracting the 2d tensor network. In this paper we get up to $D=9$ and $\chi$ of the order of hundreds. For every $D$, our results are always converged in $\chi$, so here we do not discuss the dependence on this parameter explicitly. Moreover, we also use the so called {\it simple-update} \cite{Jiang2008,Corboz2010a,Corboz2010} to find the PEPS tensors, and used a corner transfer matrix method \cite{Nishino1996,Orus2009,Corboz2010a} adapted to arbitrary unit cell sizes \cite{Corboz2014a} to compute the environment and expectation values of local operators. 

We compare various competing low-energy states in the AFHS model by using different unit cell sizes periodically repeated on the TN, e.g., $2\times2$ and $4\times4$. These unit cells correspond to $12$ and $48$ sites on the original star lattice, and reveal the true ordering of the underlying state by analyzing the bond energies as order parameter.

\section{Competing VBS Phases}
\label{sec:vbsphases}

In this section, we present our iPEPS results for the ground-states of the AFH model on the star lattice in different Heisenberg exchange couplings.

\subsection{Large-$J_e$ limit ($J_t/J_e\ll 1$)} 

 \begin{figure}
	\centerline{\includegraphics[width=\columnwidth]{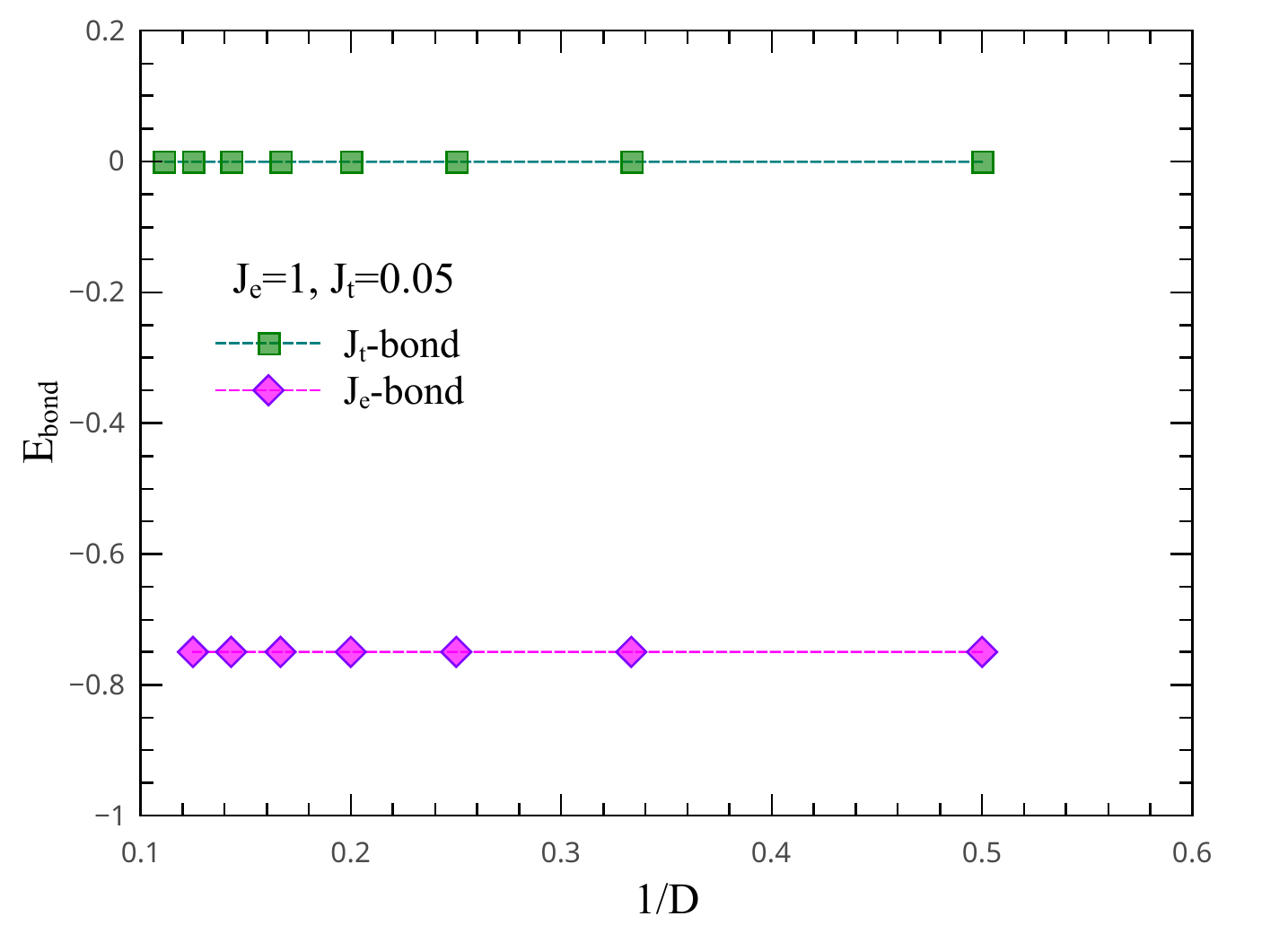}}
	\caption{(Color online) Scaling of the iPEPS bond energies with respect to inverse bond dimension $D$ in the large-$J_e$ limit. The dominating bond energies on the $J_e$ links reveal the $J_e$-dimer VBS order (Fig.~\ref{Fig:PhaseDiag}-left) in the ground state of the system.}
	\label{Fig:EbondJt0-05}
\end{figure}

We first discuss the iPEPS results obtained in the large-$J_e$ limit. In the extreme case  where $J_t=0$, the ground state of the system is a product state of isolated singlets (dimers) on the $J_e$ links, i.e., a $J_e$-dimer VBS state (see Fig.~\ref{Fig:PhaseDiag}-left). By turning on the $J_t$ couplings on the triangles, small amounts of quantum fluctuations, induced by geometric frustration, start to show up in the system. However, as long as the $J_t$ couplings are kept small, these quantum fluctuations remain small as well, in such a way that the $J_e$-dimer VBS state remains stable. Fig.~\ref{Fig:EbondJt0-05} depicts the scaling of ground-state energy versus the inverse of the bond dimension $D$ for $J_e=1, J_t=0.05$. The lowest energy states we found on $2\times2$ and $4\times4$ unit cell sizes for different bond dimensions are similar and all of them have strong bond energies on the $J_e$ links with energies $E_b^e=-0.749526$ and weak bond energies on the $J_t$ links with energies $E_b^t=-0.000475$ ($D=9$). These results reveal a $J_e$-dimer VBS order in the $D\longrightarrow\infty$ limit. Since all of the bond energies on all topologically equivalent bonds of the lattice are equivalent, this VBS state fully respects the symmetries of Hamiltonian \eqref{eq:H_AFHS} on the star lattice. Excellent energy convergence for all the values of $D$ further confirms the relatively-small amount of entanglement in the ground-state of the system. We find the ground state energy per site $\varepsilon_0=-0.375234$ for $D=9$. Let us also stress that these findings are in full agreement with previous ED and mean-field results \cite{Richter2004,Yang2010}.

\subsection{Large-$J_t$ limit ($J_t/J_e\gg 1$)} 

 \begin{figure}
	\centerline{\includegraphics[width=\columnwidth]{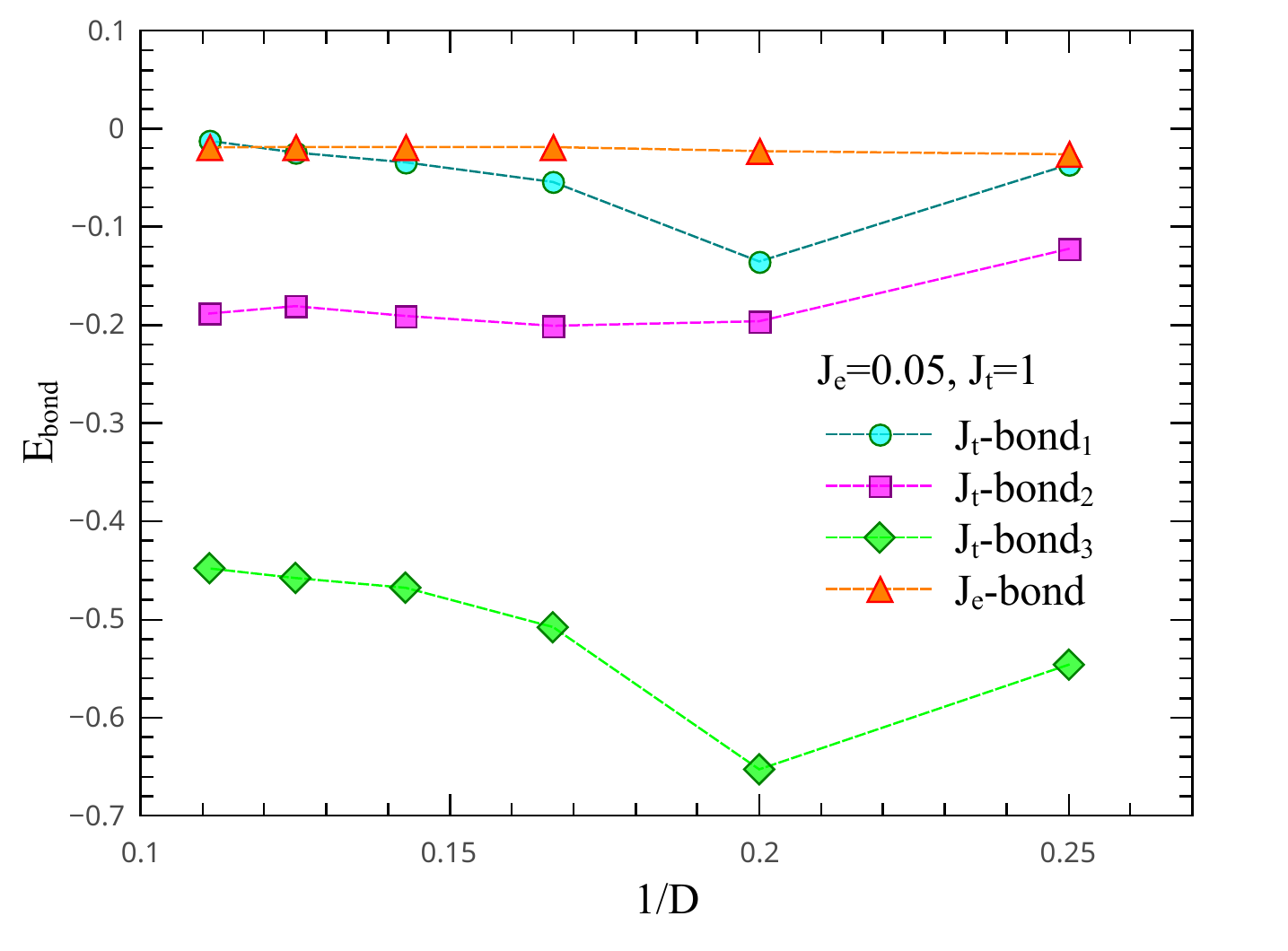}}
	\caption{(Color online) Scaling of the iPEPS bond energies with respect to inverse bond dimension $D$ in the large-$J_t$ limit for the bonds shown in the zoom panel of Fig.~\ref{Fig:PhaseDiag}. Similar results hold for other topologically equivalent bonds on the star lattice. The different magnitudes of the bond energies confirms the strong-weak bond pattern of the $J_t$-dimer VBS ground-states in the large-$J_t$ limit.}
	\label{Fig:EbondJe0-05}
\end{figure}

 \begin{figure}
	\centerline{\includegraphics[width=\columnwidth]{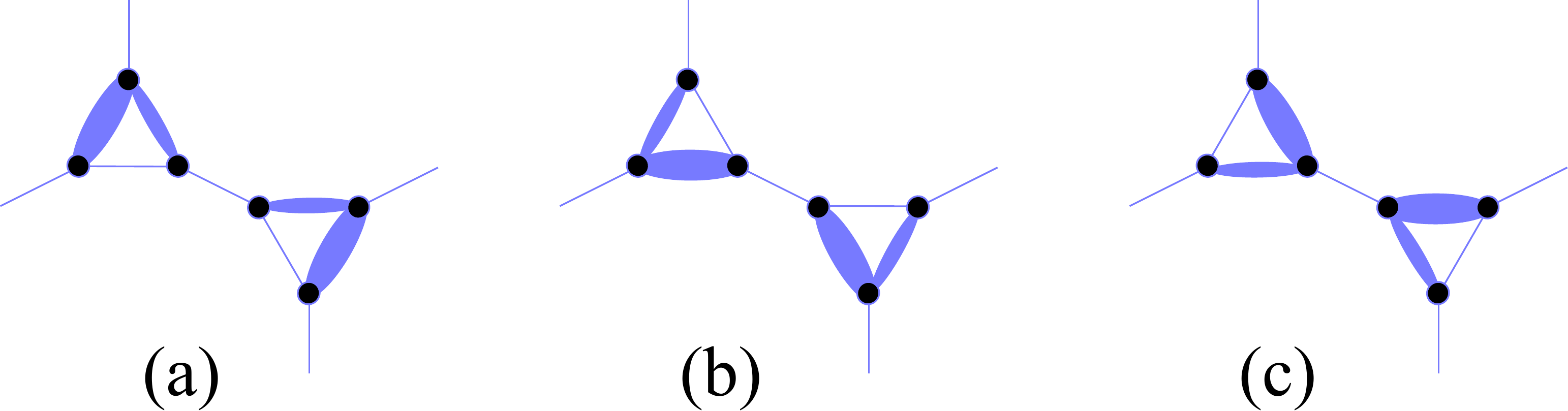}}
	\caption{(Color online) (a)-(c) Unit cells of strong-weak bond patterns with six-site order corresponding to the three degenerate $J_t$-dimer VBS ground-states of the AFHS model in the large-$J_t$ limit obtained with iPEPS. An equal superposition (resonance) of the three is also a possible ground state, but the iPEPS method favours one of them since it has lower entanglement.}
	\label{Fig:DGs}
\end{figure}

This limit is by far the most important and interesting regime of the problem. In this limit the AF couplings are much larger on $J_t$ links compared to $J_e$ bonds, and even in the extreme case where $J_e=0$, the system suffers from strong geometric frustration due to the triangles. The strong quantum fluctuations due to this frustration makes both formation and detection of quantum order in the ground state of the AFHS model a complicated task. This situation becomes even worst when the $J_e$ couplings are switched on, since more quantum fluctuations are introduced into the system. Previous studies with ED on finite-size clusters \cite{Richter2004,Misguich2007} and mean-field results \cite{Yang2010} predicted a VBS state with $\sqrt{3}\times\sqrt{3}$ order on the unit cells with 18 sites. However, it is by far not clear whether such an ordering is stable in the thermodynamic limit or not. In order to check this and find the true ordering of the ground-state, we applied the iPEPS method for fixed couplings $J_e=0.05, J_t=1$ and extracted the ground-state of the system on both $2\times2$ and $4\times4$ unit cell sizes. Let us note that the $4\times4$ unit cell consists of $48$ sites and already encompasses the $\sqrt{3}\times\sqrt{3}$ structure, and therefore is capable of detecting such an ordering, if existing, in the thermodynamic limit. 

Fig.~\ref{Fig:EbondJe0-05} illustrates the scaling of the bond energies on the $J_t$ edges of the triangles and on the $J_e$ bonds, as shown in the zoomed panel of Fig.~\ref{Fig:PhaseDiag}. As one can see, due to the large amount of quantum fluctuations and correspondingly large entanglement in the system, energy convergence at small bond dimension $D$ is rather poor. However, as we inject more entanglement into the system and go to larger $D$, the bond energies converge and reveal the true ordering of the ground state. We found the lowest bond energies $E_b^{t_1}=-0.448037$, $E_b^{t_2}=-0.188271$, $E_b^{t_3}=-0.012245$ and $E_b^e=-0.018752$ for $D=9$. Mapping these bond energies to the unit cell of the star lattice, we find a VBS state with six-site unit cell order and with a strong-weak bond pattern (see Fig.~\ref{Fig:PhaseDiag}-right) for both $2\times2$ and $4\times4$ unit cell sizes in the $D\longrightarrow\infty$ limit. Repeating the iPEPS simulations with different initial states, we found two other degenerate VBS states with the same magnitude of bond energies but different strong-week bond pattern. Putting our iPEPS results together and in perspective we find that, in contrast to previous ED \cite{Richter2004} and mean-field results \cite{Yang2010}, the true nature of the ground state of the system in the thermodynamic limit for large-$J_t$ couplings seems to be  a three-fold degenerate resonating $J_t$-dimer VBS state with six-site unit cell order. The three degenerate VBS states are constructed from repeating the unit cells of strong-week bond patterns of Fig.~\ref{Fig:DGs} on the star lattice (see Appendix.~\ref{app:en-scaling} for the full patterns). These VBS states break the $C_3$ rotational symmetry of the triangles of the star lattice, and are indeed different from the strong-weak bond patterns of the $\sqrt{3}\times\sqrt{3}$ order in Ref.~\cite{Richter2004,Yang2010}. 

The fact that similar ordering is obtained on a $2\times2$ unit cell with $12$ sites strongly confirms that the $\sqrt{3}\times\sqrt{3}$ order is not stable in the thermodynamic limit and is possibly an artifact of finite size effects and the low amounts of entanglement present on finite clusters and mean-field states. Finally, the lowest ground-state energy that we obtained obtained in this limit for $J_e=0.05, J_t=1$ is $\varepsilon_0=-0.255155$ for $D=9$. 

\subsection{Isotropic point ($J_e=J_t=1$)} 

\begin{figure}
	\centerline{\includegraphics[width=\columnwidth]{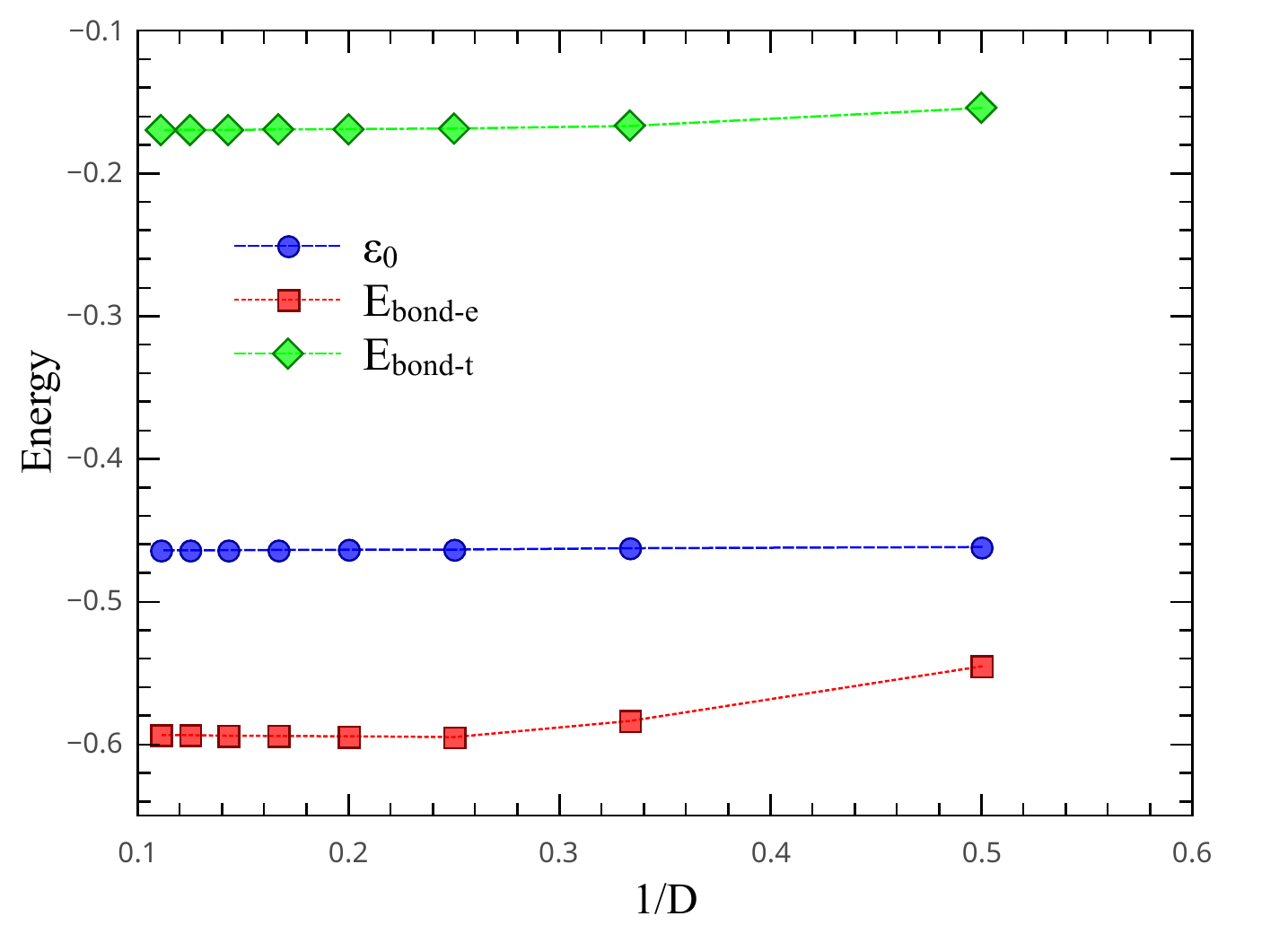}}
	\caption{(Color online)  Scaling of the iPEPS bond energies with respect to inverse bond dimension $D$ at the isotropic point, $J_e=J_t=1$. The dominating bond energies on the $J_e$ links reveal that the ground state of the system is in the same phase than the VBS state at large-$J_e$.}
	\label{Fig:EIso}
\end{figure}

Next we focus on the isotropic case where $J_e=J_t=1$. It has already been shown that the classical ground states of the AFHS model at this point \cite{Richter2004} are similar to those of the kagome lattice \cite{Richter2004a} and are described by two candidate states, i.e., a $\sqrt{3}\times\sqrt{3}$ and a $q=0$ state. By introducing quantum fluctuations over these classical ground-states by means of linear spin-wave theory (LSW), the classical order is destroyed and the system is described by uniform bonds with energy $E_b^{LSW}=-0.296759$. Further analysis by ED on finite clusters also predicted a singlet ground state with bond energy $E_b^{ED}=-0.309918$ for $N=42$ lattice sites \cite{Richter2004}. 

In order to find the true ordering of the ground-state in the thermodynamic limit, we applied the iPEPS method to the AFHS model on different unit cell sizes and different bond dimensions $D$. We find that, in contrast to the ED and LSW results \cite{Richter2004}, the iPEPS ground state in the thermodynamic limit is described by a VBS state with strong bond energies on $J_e$ links with $E_b^{e}=-0.593559$ and weak bond energies on $J_t$ links with $E_b^{t}=-0.169517$ for $D=9$ (see Fig.~\ref{Fig:PhaseDiag}-middle). This is nothing but another $J_e$-dimer VBS state with similar properties to the VBS phase in large-$J_e$ limit. 

Fig.~\ref{Fig:EIso} presents results for the bond energies and ground state energy per site for the isotropic pint obtained with setup $A$. The excellent convergence of energies with respect to the inverse bond dimension $D$ shows a strong (weak) pattern on the $J_e$ ($J_t$) links and further confirms that the ground state of the AFHS model at this point is a $J_e$-dimer VBS state.

Later we will show that, in fact, this state is in the same phase as the VBS state at large-$J_e$ limit, so that it has the same type of order but with a larger correlation length since it is closer to the quantum critical point. 

We have also calculated the ground-state energy per site for the AFHS model at the isotropic point in the infinite $D$ limit, $\varepsilon_0=-0.464114$, which is in good agreement with the predicted ED energy $\varepsilon_0^{ED}=-0.464877$.

\section{Two-point correlators}
\label{sec:correlators}

\begin{figure}
	\centerline{\includegraphics[width=\columnwidth]{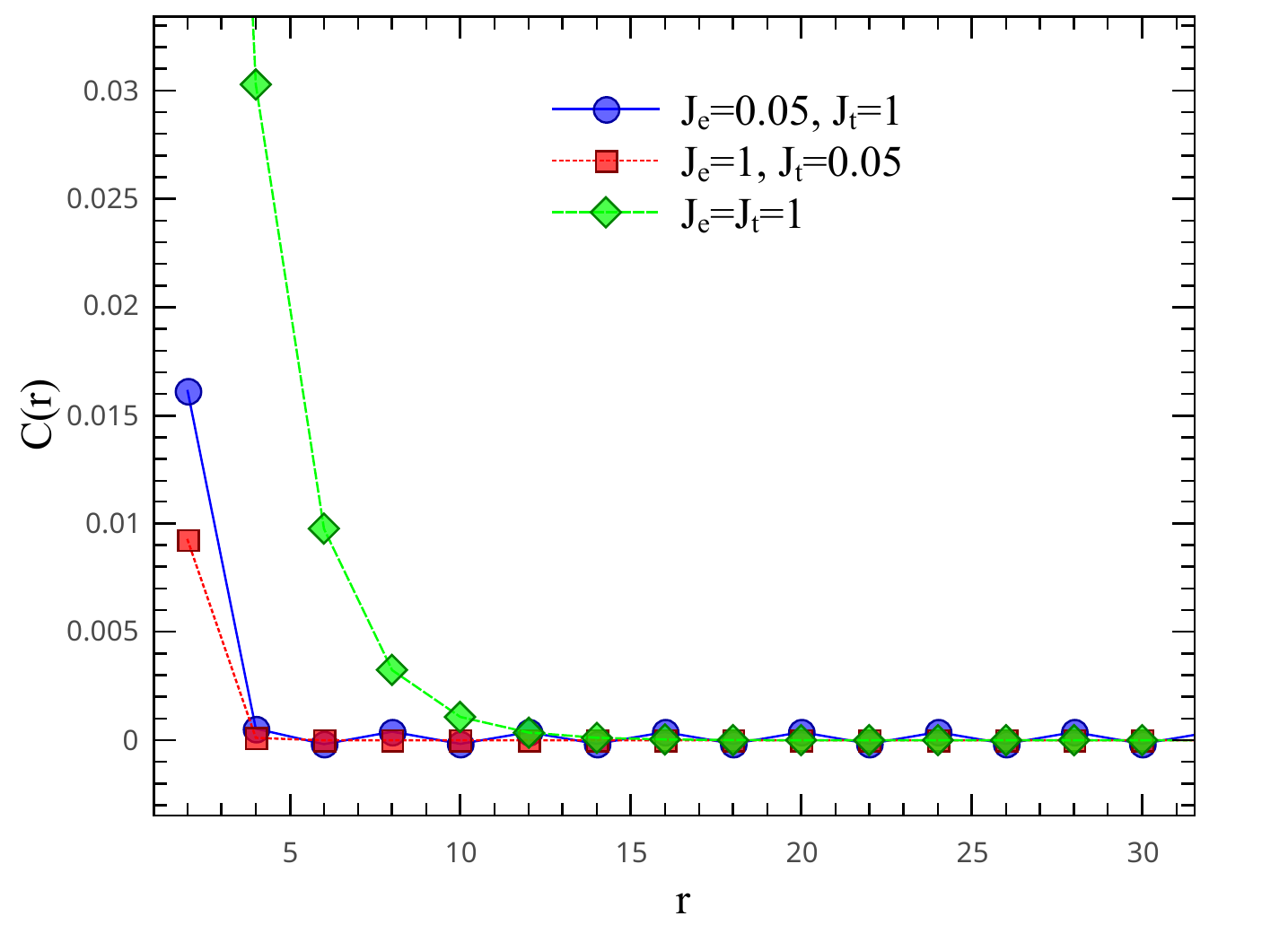}}
	\caption{(Color online) Spin-spin correlator $C(r)=\expectval{\mathbf{S}_{(x,y)} \mathbf{S}_{(x,y+r)}}$ for the three limiting cases of the AFHS model, i.e., large-$J_e$ limit (red), large-$J_t$ limit  (blue) and the isotropic point (green).}
	\label{Fig:Cr}
\end{figure}

In order to provide more insight regarding the nature of the VBS states of the AFHS model in different regimes, we also calculated the spin-spin correlator $C(r)=\expectval{\mathbf{S}_{(x,y)} \mathbf{S}_{(x,y+r)}}$ for the three limiting cases, i.e., large-$J_e$ limit, large-$J_t$ limit and the isotropic point. In Fig.~\ref{Fig:Cr} we plot this correlator for these three cases as obtained from iPEPS ground states with $D=9$.

\begin{figure*}
	\centering
	\begin{tabular}{cc}
		\includegraphics[width=9cm]{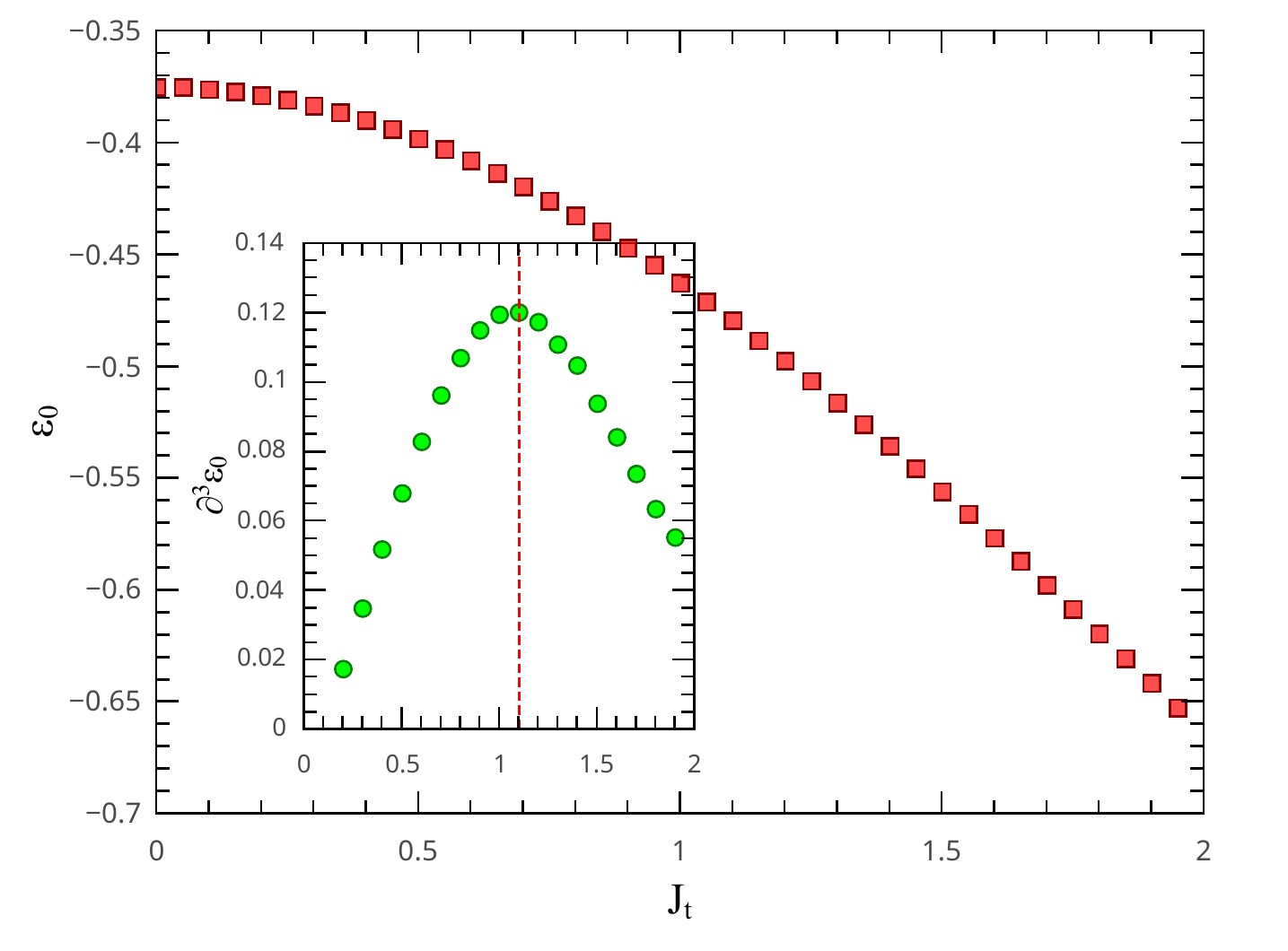} & \includegraphics[width=9cm]{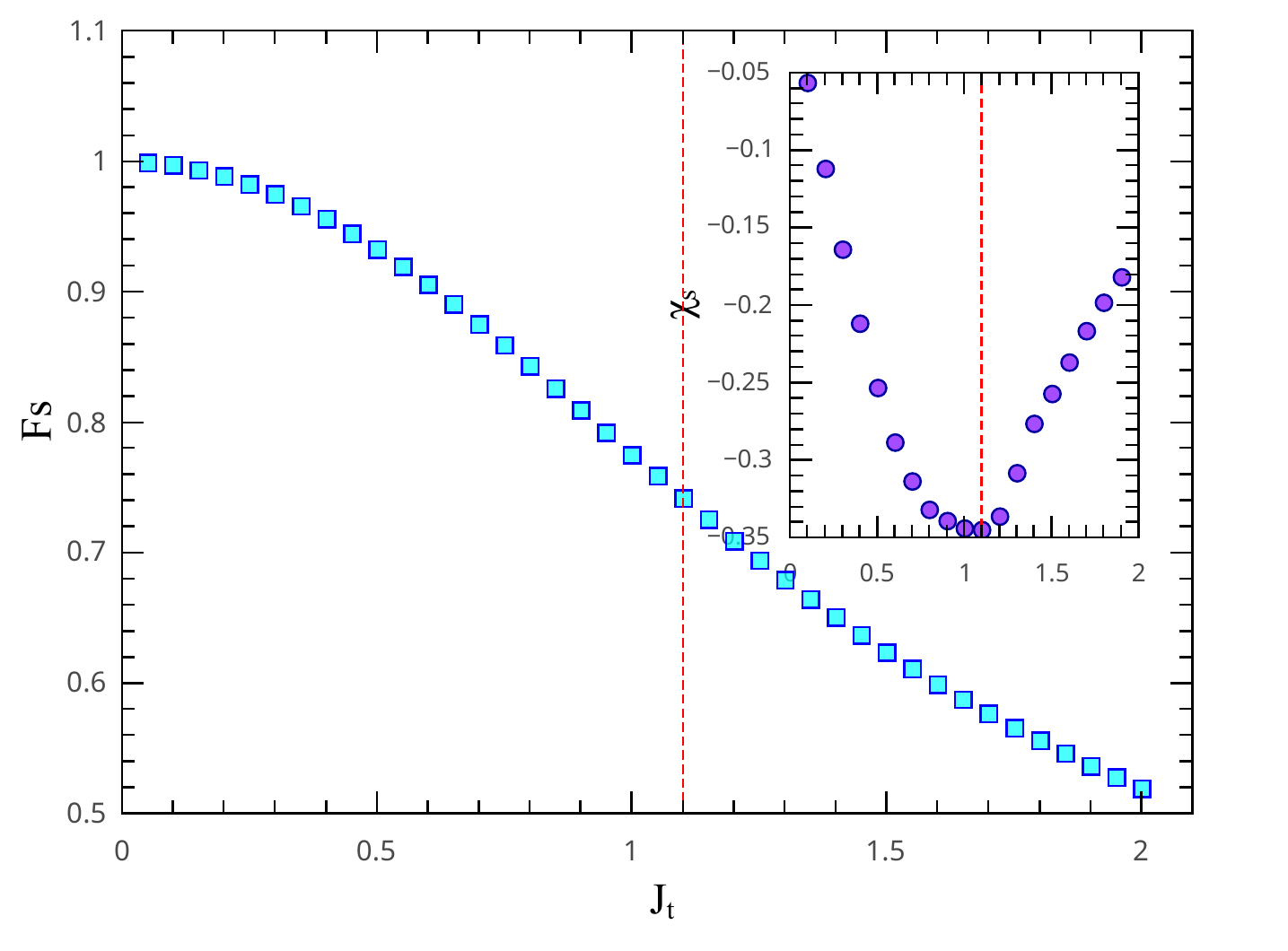} \\
		(a) & (b) 
	\end{tabular}
	\caption{(Color online) (a) Ground state energy per site of the AFHS model for $J_e=1$ and $0\leqslant J_t\leqslant 2$ obtained from iPEPS with bond dimension $D=9$. The inset shows the third-derivative of the energy with respect to $J_t$. A QPT is detected from the derivative of the energy at $J^c=J_t/J_e \approx 1.1$. (b) Ground state fidelity per site, Eq.~\eqref{eq:fidelity}, from the iPEPS ground-states on a $4\times 4$ unit cell for $D=9$. The inset shows the fidelity susceptibility $\chi_s$.}
	\label{Fig:E0-Fid} 
\end{figure*}

We find that the correlations for VBS states at the two extreme limits, i.e., the $J_e$-dimer VBS state (red curve) and the resonating $J_t$-dimer VBS states (blue curve), decay both exponentially fast. The correlation for these VBS states is therefore short range and spreads only between the two neighboring spins at most. We found that this behavior did not change substantially when increasing the bond dimension of the iPEPS. However, the correlation at the isotropic point (green line) spreads to larger distances, almost up to $10$ lattice sites farther. This is explained by noting that the isotropic point is very close to the critical point at  $J^c=J_t/J_e \approx 1.1$ (see Sec.~\ref{sec:qpt}). As approaching the critical point, entanglement diverges in the system and correlation becomes long range. The spin-spin correlation, therefore, decays slower to larger distances.We also observe that this correlator is more sensitive to increasing the bond dimension $D$, since it is closer to a quantum critical point. 

\section{Quantum phase transition} 
\label{sec:qpt}

We already showed that the AFHS model hosts two different VBS states at the two extreme limits of the AF couplings. It is therefore natural to expect a quantum phase transition (QPT) between the two limiting cases by varying the AF exchange couplings. In order to reveal the phase transition and to extract the phase diagram of the model in the thermodynamic limit, we fixed $J_e=1$ and studied the ground state of the system and its energy for $0\leqslant J_t\leqslant 2$. We used derivatives of the ground state energy as well as the ground state fidelity to precisely pinpoint the critical point and nature of the QPT.

Fig.~\ref{Fig:E0-Fid}-(a) shows the ground state energy per site for $D=9$. The inset further shows the third-derivative of the energy with respect to $J_t$ for the whole regime of AF parameters. One can clearly see that a QPT is best detected at $J^c=J_t/J_e \approx 1.1$, which is close to the predicted ED value $J^c_{ED}=1.3$ \cite{Misguich2007,Yang2010}. This slight difference can be another artifact of finite size effects in ED calculations. Let us further note that the smooth energy curve and the fact that the critical point is only captured with second and higher derivatives of the energy clarifies that the QPT is continuous. 

The location of critical point is further confirmed with the ground state fidelity, Eq.~\eqref{eq:fidelity}, of the iPEPS ground-states and the derivative of fidelity, i.e., the fidelity susceptibility. Fig.~\ref{Fig:E0-Fid}-(b) shows the fidelity of the AFHS model obtained on a $4\times 4$ iPEPS unit cell for $J_e=1$ and $0\leqslant J_t\leqslant 2$. $\ket{\Psi(\lambda_1)}$ corresponds the ground state at $J_e=1, J_t=0$ and $\ket{\Psi(\lambda_2)}$ corresponds to different $J_t$ at each step. The QPT is best detected  at $J^c=J_t/J_e \approx 1.1$ by the fidelity susceptibility, $\chi_s$, at the inset of Fig.~\ref{Fig:E0-Fid}-(b). The smooth variation of the fidelity per site further confirms the continuous nature of the QPT \cite{Zhou2008,Jahromi2018} in the AFHS model.

\begin{figure}
	\centerline{\includegraphics[width=\columnwidth]{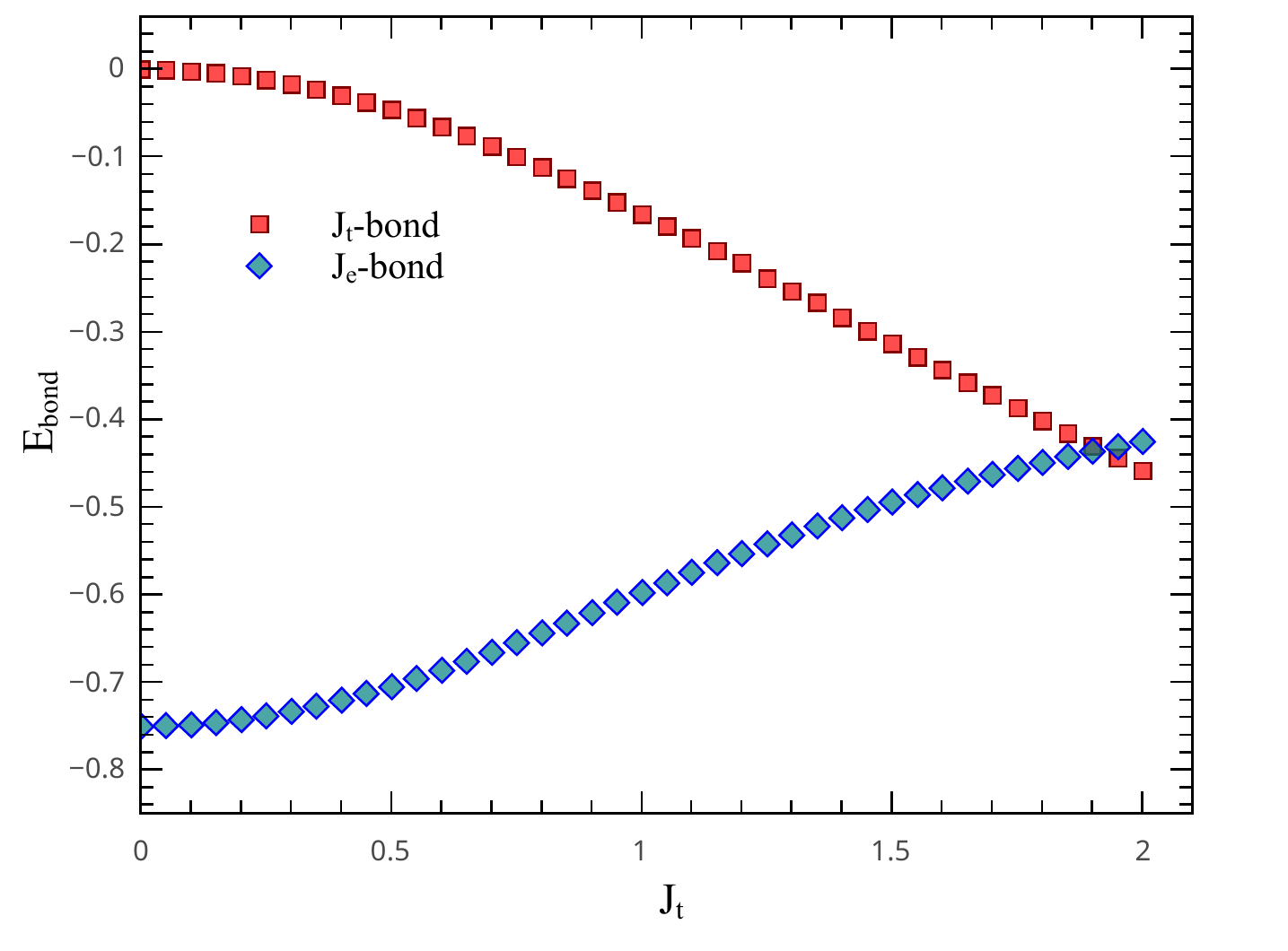}}
	\caption{(Color online) Bond energies on $J_t$ and $J_e$ links considered as order parameters.}
	\label{Fig:OP}
\end{figure}

The fidelity results are further in agreement with the bond energies, which we consider as order parameter. Fig.~\ref{Fig:OP} illustrates the bond energies on $J_e$ and $J_t$ links as a function of $J_t$. As we already noted, the VBS state at large-$J_e$ couplings is a VBS state with strong energies on $J_e$ bonds and almost zero energies on $J_t$ bonds. By increasing the $J_t$ couplings, the bond energies on the $J_e$ dimers start to diminish and the VBS state on the $J_t$ links gradually appears. This behaviour is best seen in Fig.~\ref{Fig:OP} which clearly shows the energy increase (decrease) on $J_t$ ($J_e$) links when increasing the $J_t$ couplings.

Finally, we emphasize how close the isotropic point is to the critical point. This shows why the ground state of the system at isotropic point is in the $J_e$-dimer VBS phase, but with larger correlation length.

\section{Conclusions and outlook} 
\label{sec:conclude}

In this paper we have studied the spin-$1/2$ antiferromagnetic Heisenberg model on the star lattice using tensor network methods based on the iPEPS algorithm adapted for arbitrary size unit cells on triangle-honeycomb lattices. We found that the ground states of the AFHS model host two competing VBS states in different AF Heisenberg exchange couplings, i.e., a VBS state with strong dimers on the expanding bonds of the lattice, and which fully respects the symmetries of the model, and a three-fold degenerate resonating VBS state with six-site unit cell order, and with patterns of strong-weak dimers on the triangles that break $C_3$ symmetry. We found that, in contrast to previous ED and mean-field results, the $\sqrt{3}\times\sqrt{3}$ order is not the stable ordering of the ground state at the thermodynamic limit in the large-$J_t$ regime. Moreover, we found that the ground-state of the system at the isotropic point with uniform AF bonds on the star lattice is, in contrast to previous findings, a VBS state with strong (weak) bond energies on the expanded (triangle) links of the star lattice. We also studied the quantum phase transition in the system and computed the zero-temperature phase diagram in the thermodynamic limit. We located a continuous QPT at $J^c=J_t/J_e \approx 1.1$ by calculating the energy derivatives and ground state fidelity using  large unit cells.

Our findings and explored phases may be realized in recently discovered polymeric Iron Acetate. It has already been shown that \cite{Zheng2007} although the Iron Acetates eventually orders magnetically at low temperatures, the magnetic ordering temperature is much lower than the estimated Curie-Weiss temperature, revealing the frustrated nature of the spin interactions. Our finding for frustrated spin-$1/2$ Heisenberg model on the star lattice can shed light on the physics of frustration in such systems and provide more insight for future experiments. Moreover, our iPEPS algorithm developed for the star lattice can also be used for further investigations of the model in the presence of a magnetic field,  to capture possible magnetization plateaus and their underlying exotic phases, which we leave for future works. 
 
\section{Acknowledgements}
S.S.J. acknowledges the support from Iran Science Elites Federation (ISEF). The iPEPS calculations were performed on the HPC cluster at Sharif University of Technology and the MOGON cluster at Johannes Gutenberg University.

\appendix

\section{${\rm i}$PEPS for the star lattice}
\label{app:ipeps}  

The infinite Projected Entangled-Pair State method \cite{Verstraete2004,Verstraete2006,Vidal2007} is a variational Tensor Network ansatz for approximating the grouund state of 2d quantum systems in the thermodynamic limit. It can be considered as a two-dimensional generalization of the Matrix Product State (MPS) \cite{Fannes1992,Ostlund1995}, which is the heart of 1s TN methods such as DMRG \cite{White1992,White1993} and iTEBD \cite{Vidal2007,Orus2008}. The most advanced and efficient iPEPS algorithms are typically developed for the square lattice, where the ansatz is composed of unit cells of rank-$5$ tensors, periodically repeated on the lattice \cite{Corboz2010,Corboz2010a}. In this context, the contraction of the infinite 2d TN required for calculating the norm and expectation values is performed via approximation methods such as the Corner Transfer Matrix (CTM) Renormalization Group \cite{Nishino1996,Orus2009}. 

\begin{figure}[h]
	\centerline{\includegraphics[width=\columnwidth]{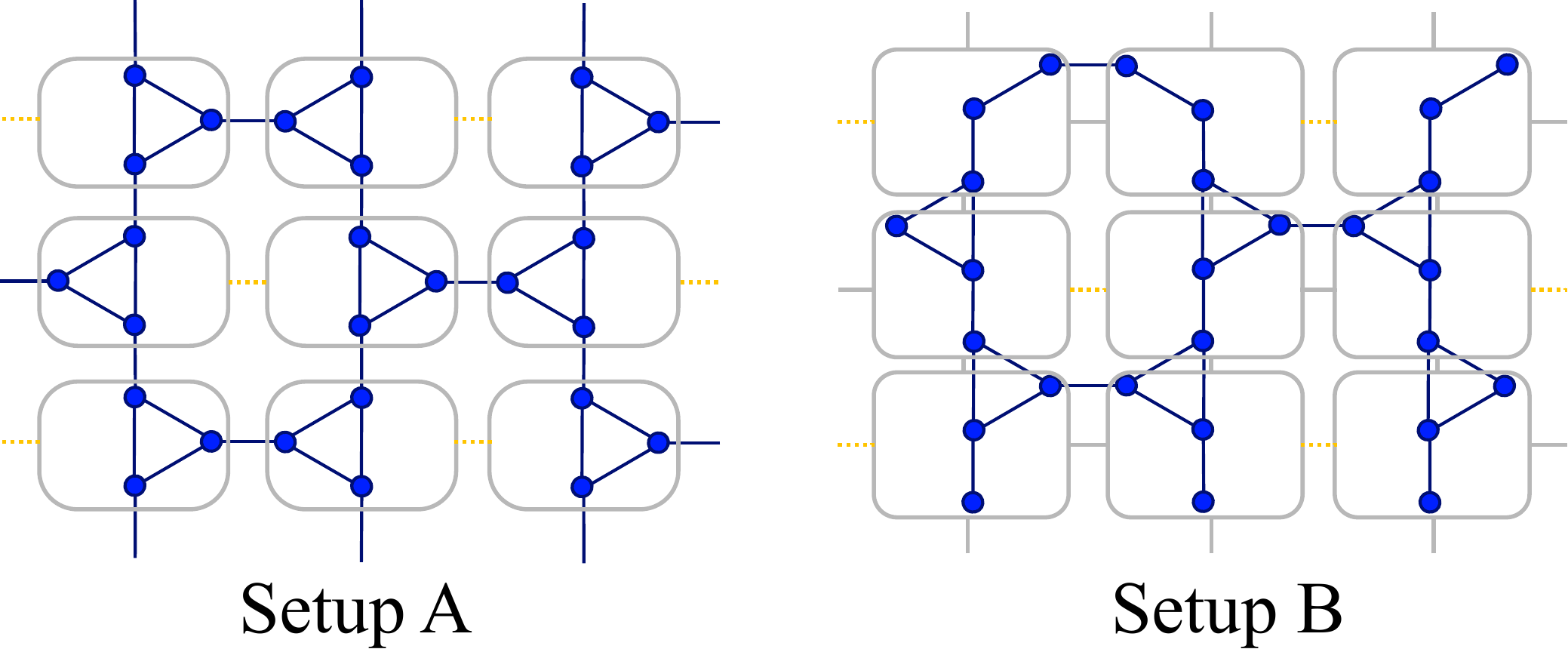}}
	\caption{(Color online) Two possible iPEPS for the star lattice designed by different strategies for constructing the block sites. In Setup $A$, the three spins at the vertices of triangles are grouped together and form a block site with local Hilbert space dimension $d=2^3$. In Setup $B$, the three spins belonging to two edges of a triangle and a $J_e$ bond are grouped together. Next, we associate a tensor to each block site and introduce trivial bonds with dimension $D=1$ (yellow dotted links) along the directions with no interaction in order to build the square-larttice TN. See also Ref.~\cite{Jahromi2018} for more details.}
	\label{Fig:IPEPSSetup}
\end{figure}

In Ref.~\cite{Jahromi2018}, we developed an improved version of the iPEPS algorithm, and explained how to use the current state-of-the-art iPEPS methods for triangle-honeycomb lattices such as star lattice. The basic idea behind this methodology is to map the star lattice to a brick-wall honeycomb lattice of coarse-grained block sites with local Hilbert space dimension $d=2^3$. Next, by associating an iPEPS tensor to each block site, and introducing trivial ``dummy"  indices with bond dimension $D=1$ (as in the yellow dotted lines in Fig.~\ref{Fig:IPEPSSetup}), the iPEPS TN on the square lattice is obtained. 

In order to capture different VBS orders for different exchange couplings in the antiferromagnetic Heisenberg model on star lattice, we introduced two different setups for grouping the three spin-$1/2$ degrees of freedoms into block sites (Setups $A$ and $B$ in Fig.~\ref{Fig:IPEPSSetup}). In Setup $A$, which we used for large-$J_e$ limit, the three spins at the vertices of the triangles are grouped into block sites and $J_e$ links coincide on the virtual ledges of the iPEPS tensors. This choice of grouping the vertices leads to a bias towards a trimer state on the triangles. This effectively corresponds to having an infinite $D$ between the sites within a block; i.e., all correlations between the sites in a block are taken into account exactly. However, as we will see in the next section, since the $J_t$ couplings are week in the large-$J_e$ regime, this bias will vanish in the large $D$ limit and the true ordering is retrieved in the thermodynamic limit.

As outlined in the main text, the large-$J_t$ limit suffers from large geometric frustration and correspondingly, large quantum fluctuations are present in the system. Therefore, biasing the iPEPS tensors towards trimerized states on the triangles might hamper the convergence of the algorithm to the true ground states. We therefore choose another strategy for grouping the vertices of the lattice. In this respect, we group the three sites belonging to two edges of a triangle and one $J_e$ link into a block site as shown in Fig.~\ref{Fig:IPEPSSetup}. This choice of block sites helps in the  convergence of the iPEPS algorithm to the the states with lower energies in the large-$J_t$ regime.

Let us further remark that we use an improved version of the CTM algorithm designed for arbitrary unit cell sizes \cite{Corboz2014a} to approximate the environment and calculate the norm and expectation values of local operators.

\section{Ground state fidelity for n-site unit-cells from iPEPS and CTMs}
\label{app:fid}  

\begin{figure}
	\centerline{\includegraphics[width=\columnwidth]{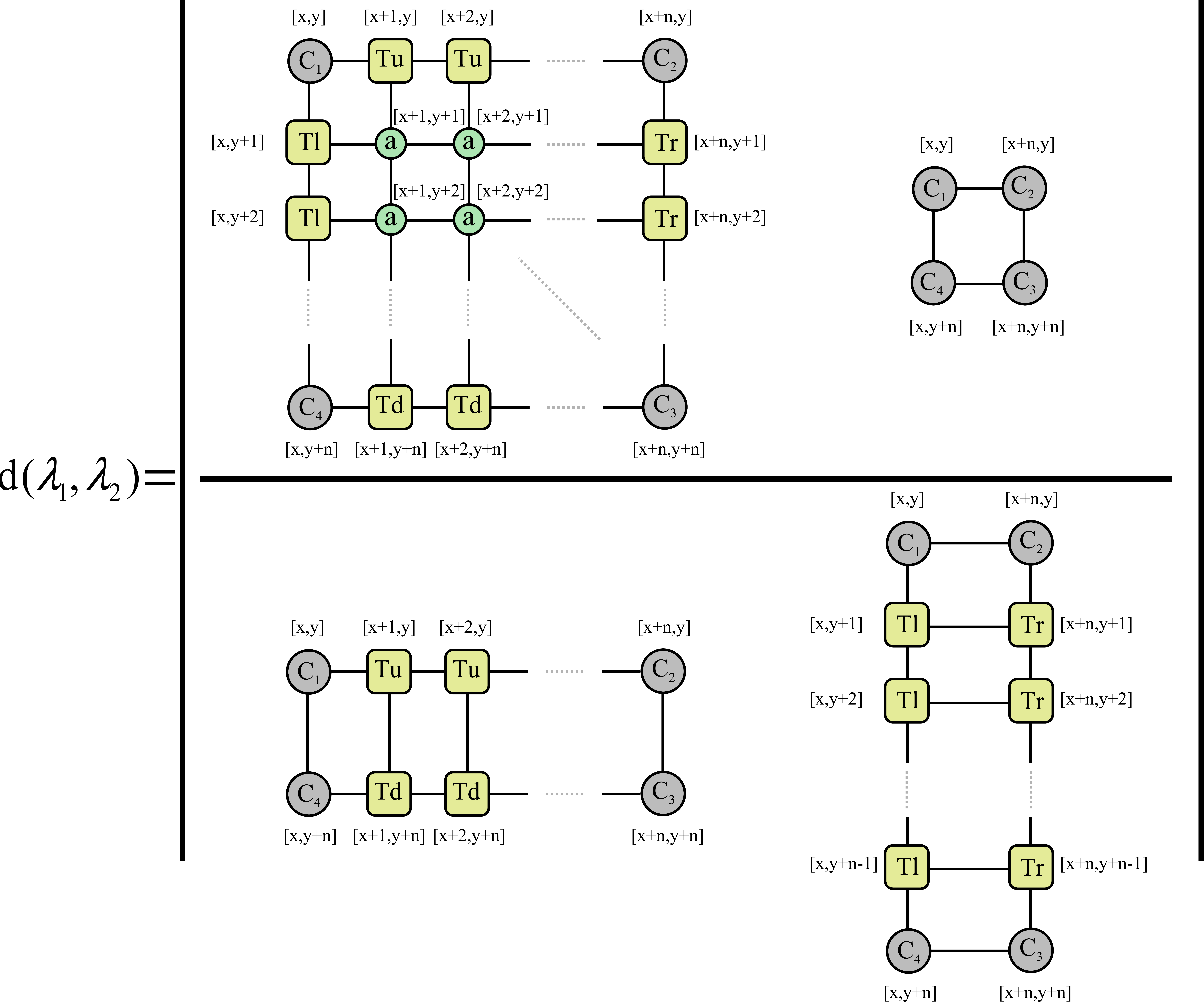}}
	\caption{(Color online) Contraction providing $d(\lambda_1,\lambda_2)$, expressed in terms of the four corner tensors $C_1, C_2, C_3,C_4$ and the half-row and half-column transfer matrices $T_u, T_r, T_d$ and $T_l$. Indices $[x,y]$ correspond to the relative position of tensors in the network with respect to the unit cell. $a[x,y]$ are double (reduced) tensors with bond dimension $D^2$, constructed form contracting rank-$5$ PEPS tensors  $A(\lambda_1)$ and $A(\lambda_2)$ along their physical dimensions $d$.} 
	\label{Fig:fid}
\end{figure}

In this section, we provide the details for calculating the ground state fidelity for iPEPS unit cells with arbitrary sizes, by using the CTM method.

Consider a quantum lattice system with Hamiltonian $H(\lambda)$, $\lambda$ being a control parameter. For two different values $\lambda_1$ and $\lambda_2$ of this control parameter, we have ground-states $\ket{\Psi (\lambda_1)}$ and $\ket{\Psi(\lambda_2)}$. The ground-state fidelity is then given by $F(\lambda_1, \lambda_2) = |\braket{\Psi(\lambda_2)}{\Psi(\lambda_1)}|$, which scales as $F(\lambda_1, \lambda_2) \sim d(\lambda_1, \lambda_2)^N$, with $N$ the number of lattice sites. One therefore defines the \emph{fidelity per lattice site} as \cite{Zhou2008}
\be
\ln d(\lambda_1, \lambda_2) \equiv \lim_{N \rightarrow \infty} \frac{\ln F(\lambda_1, \lambda_2)}{N}. 
\ee
Using the CTM language, we showed in Ref.~\cite{Jahromi2018} that the fidelity per site is given by
\be
d(\lambda_1, \lambda_2) =\left|Ê \frac{\bra{\Sigma_U} E \ket{\Sigma_D} \braket{\Omega_U}{\Omega_D}}{{\braket{\Sigma_U}{\Sigma_D}\bra{\Omega_U} K \ket{\Omega_D}}} \right|,
\label{eq:fidelity}
\ee
where the terms in the numerator and denominator correspond to the overlap between the dominant eigenvectors of the infinite 1d transfer matrix describing the fidelity between two ground states. A graphical representation of Eq.~\eqref{eq:fidelity} for TNs with one-site unit cell has already been given in Ref.~\cite{Jahromi2018}. In Fig.~\ref{Fig:fid} we provide the details for calculating the fidelity per site, $d(\lambda_1, \lambda_2)$, for two ground states composed of multiple tensors in a unit cell with an arbitrary number of sites (tensors). Further details about the tensors are provided in the caption of the figure.

Independently of the nature of the underlying phases, the ground state fidelity is a powerful probe for capturing quantum phase transitions (QPT) and determining their nature ~\cite{Zhou2008,Jahromi2018}.

\section{Finite-entanglement scaling of energies}
\label{app:en-scaling}  
Here we show some results for the scaling of the ground state energy with the bond dimension, as obtained with setups $A,B$ for different limiting cases of the AFHS model in the thermodynamic limit.

\begin{figure*}
	\centering
	\begin{tabular}{cc}
		\includegraphics[width=9cm]{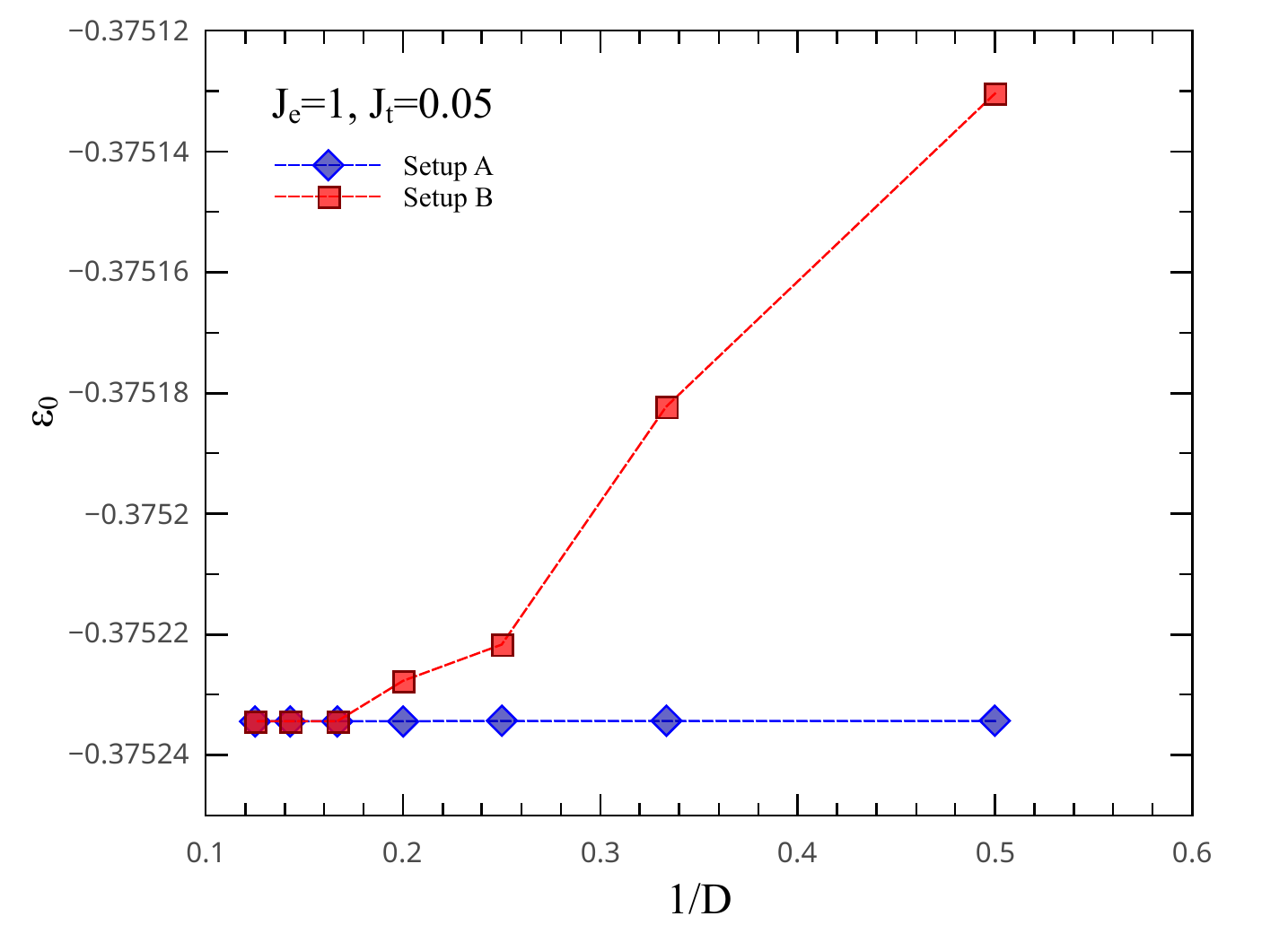} & \includegraphics[width=9cm]{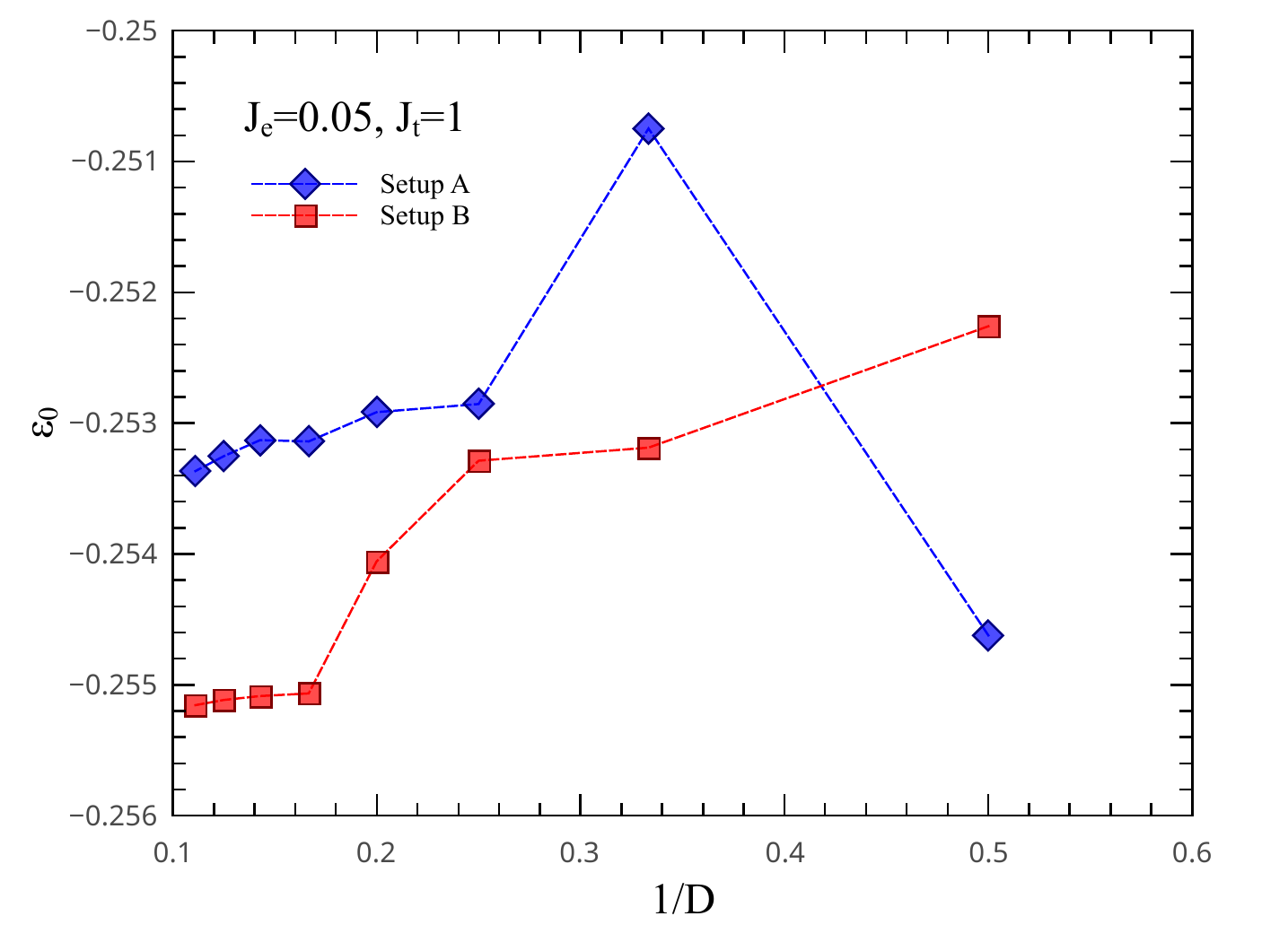} \\
		(a) & (b) 
	\end{tabular}
	\caption{(Color online) Scaling of the ground state energy per site, $\varepsilon_0$, versus inverse bond dimension $D$ (a) in the large-$J_e$ and (b) in the large-$J_t$ limits, for both setups $A$, $B$ defined in Fig.~\ref{Fig:IPEPSSetup}. As one can clearly see in (a), the convergence of energy in Setup $A$ is way much better than in Setup $B$ for large-$J_e$ couplings. However, at large $D$ both algorithms converge to the same values. On the other hand, Setup $B$ produces lower energies even at large bond dimensions $D$ in the large-$J_t$ limit in (b), and is therefore better suited for detecting the VBS ordering at large-$J_t$ couplings.}
	\label{Fig:E0} 
\end{figure*}

As outlined in previous sections, two different strategies were used to construct the block sites, and correspondingly, the TN of the star lattice. Fig.~\ref{Fig:E0}-(a) illustrates the scaling of the ground state energy per site, $\varepsilon_0$, versus inverse bond dimension $D$ in the large-$J_e$ limit for both $A$ and $B$ setups. As one can see, Setup $A$ produces better convergence particularly at small $D$. However in the large $D$ limit both algorithms converge to the same values. 

On the other hand, the convergence of setup $B$ in Fig.~\ref{Fig:E0}-(b) is much better than setup $A$ in the large large-$J_t$ limit, and therefore the algorithm produces lower energies with setup $B$. One can further see that due to the large amount of entanglement present in the large-$J_t$ regime, induced by large geometric frustration, the convergence of the algorithm is rather poor at small $D$ and the true ordering of the ground state is only captured for large $D$. This once again confirms why previous mean-field and ED results on finite-size clusters found incorrect ordering for the VBS state of the AFHS model in this regime. Let us further remind that setup $A$ is initially biased towards a trimerized state, which is far from the true ordering of the ground state in the large-$J_t$ regime. The algorithm, therefore, converges to local minima and the true ordering is not retrieved. 

\begin{figure}
	\centerline{\includegraphics[width=\columnwidth]{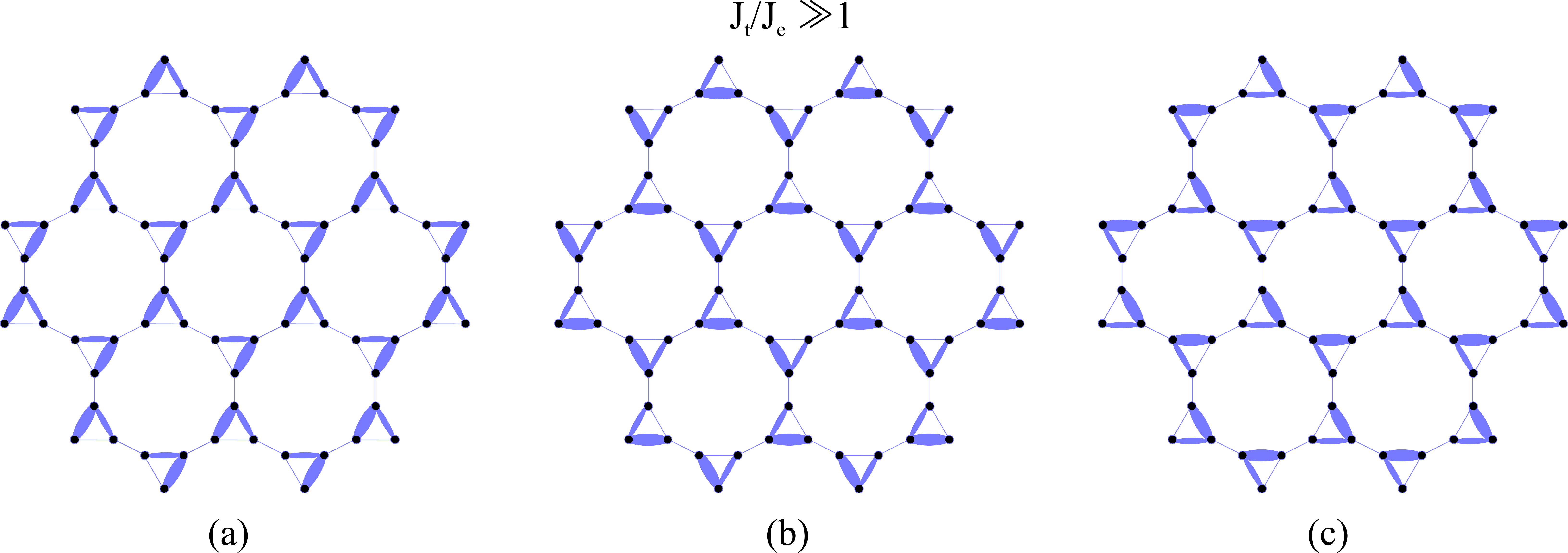}}
	\caption{(Color online) (a)-(c) The three degenerate $J_t$-dimer VBS states obtained with iPEPS in the large-$J_t$ regime. Each VBS state is obtained by applying appropriate $C_3$ rotations to the up and down triangles of the two other VBS states.}
	\label{Fig:GrounStates}
\end{figure}

By initializing the iPEPS algorithm with different random states, we obtained three degenerate $J_t$-dimer VBS states for the ground-state of the AFHS model in the large-$J_t$ regime. The three VBS states are composed of different strong-weak bond energy patterns, which are illustrated in Fig.~\ref{Fig:GrounStates}-(a)-(c). One can construct the strong-weak pattern of each VBS state by applying appropriate $C_3$ rotations to the up and down triangles of the two other VBS states. Further details regarding this phase are provided in the main text.

\bibliography{references}{}
\bibliographystyle{apsrev4-1}

\end{document}